\documentclass{article}

\usepackage{graphicx}

\def\abstract#1{\vskip 7mm 
        \begin{center}{\large Abstract}\par \smallskip
                \begin{minipage}[c]{12cm}
                        \small #1
                \end{minipage}
        \end{center}
}
\def\title#1{\begin{center}{\Large\bf #1}\end{center}}
\def\author#1{\vskip 5mm \begin{center}{#1}\end{center}}
\def\address#1{\begin{center}{\it #1}\end{center}}
\makeatletter
\@ifundefined{alt}{\def\alt{\mathrel{\mathpalette\vereq<}}}{}
\@ifundefined{agt}{}{}
\def\vereq#1#2{\lower3pt\vbox{\baselineskip1.5pt \lineskip1.5pt
\ialign{$\m@th#1\hfill##\hfil$\crcr#2\crcr\sim\crcr}}}
\makeatother

\begin{document}

\title{%
Self-Similar Solutions, Critical Behavior and 
Convergence to Attractor in Gravitational Collapse
}
\author{%
  Tomohiro Harada,\footnote{E-mail:harada@gravity.phys.waseda.ac.jp}
}
\address{%
  Department of Physics,
  Waseda University \\
  Shinjuku, Tokyo 169-8555, Japan
}

\abstract{
General relativity as well as Newtonian gravity 
admits self-similar solutions
due to its scale-invariance.
This is a review 
on these self-similar solutions and 
their relevance to gravitational collapse.
In particular, our attention is mainly paid on
the crucial role of self-similar solutions
in the critical behavior and 
attraction in gravitational collapse.  
}

\section{Introduction}
\label{sec:introduction}
The basic equation of general relativity is
the Einstein equations.
Since the Einstein equations are simultaneous second-order 
partially differential equations (PDEs),
it is not easy to obtain solutions.
It is useful to assume symmetry such as 
spherical symmetry and staticity
in obtaining solutions.
However, it is often difficult to obtain inhomogeneous
and dynamical solutions.
In this context,
it is quite useful to assume {\em self-similarity}.
For example, in spherical symmetry, 
this assumption reduces the PDEs to a set of 
ordinarily differential equations (ODEs).
Therefore, the assumption of self-similarity is 
very powerful in finding exact and/or numerical 
dynamical and inhomogeneous solutions. 
The application of self-similar solutions 
is very large including cosmological perturbations,
star formation,
gravitational collapse, primordial black holes,
cosmological voids and cosmic censorship.
Among them, the most widely researched model is 
the spherically symmetric self-similar system with perfect fluids,
which was pioneered by by Cahill and Taub (1971) in general relativity
and by Penston (1969) and Larson (1969) in Newtonian gravity.

In recent progress in general relativity,
self-similar solutions have called attention 
not only because they are easy to obtain 
but also because they may play important roles 
in cosmological situations and/or gravitational collapse.
One of the important roles of self-similar solutions 
is to describe asymptotic behaviors of more 
general non-self-similar solutions.
This role has been shown explicitly in several models
such as homogeneous cosmological models. 
In cosmological context, the suggestion that 
spherically symmetric fluctuations might naturally evolve
from complex initial conditions via the Einstein equations
to a self-similar form has been termed the `similarity
hypothesis' (Carr 1993, Carr and Coley 1999).
The hypothesis has been also suggested to hold in more 
general situations, including gravitational collapse.

In the present article, we pay attention to a 
spherically symmetric system with a perfect fluid.
For this system,
recent numerical simulations of PDEs have revealed 
two interesting behaviors in gravitational collapse, one is 
the critical behavior and the other is 
the convergence to attractor.
The critical behavior was observed around the threshold of black hole 
formation. There is a self-similar solution that sits at the 
threshold, which is called a critical solution.
For near-critical collapse, the solution first
approach the critical solution but diverges away from 
this solution after that.
There is observed the scaling law for the formed black hole
for supercritical collapse. 
This is observed only as a result of fine-tuning
the parameter which parametrizes initial data.
On the other hand, the convergence to attractor is 
observed without fine-tuning.
The density and velocity fields around the center 
approach those of another 
self-similar solution in an approach to the formation of 
singularity at the center, which is called an attractor solution.

The main claim of this review article
is that the critical behavior and 
convergence to attractor in gravitational collapse 
are both well understood in the context of self-similarity 
hypothesis and stability analysis in gravitational collapse.
The difference between these two behaviors is 
in the stability of the pertinent self-similar solution.
We also briefly 
review the recent work on how these two behaviors are 
in Newtonian gravity.

\section{Self-similarity}
\label{sec:selfsimilarity}
\begin{figure}[htbp]
\begin{center}
\includegraphics[height=7cm,width=7cm]{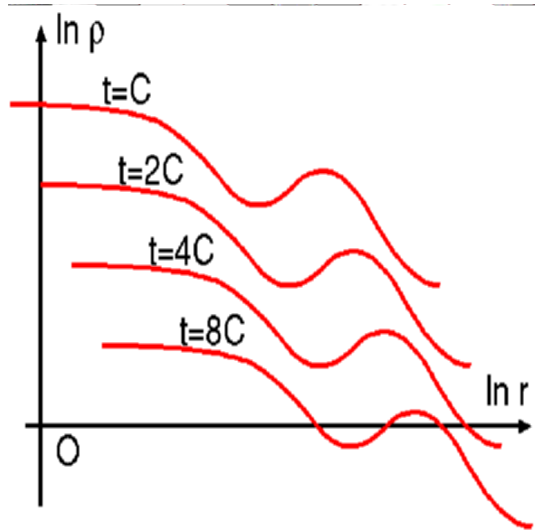}
\caption{\label{fg:similarity}
Schematic picture for the self-similar evolution of the density field.} 
\end{center}
\end{figure}

In general relativity, self-similarity or 
{\em homothety} is covariantly 
defined in terms of the homothetic vector $\xi$ which satisfies
the homothetic Killing equation
\begin{equation}
{\cal L}_{\xi}g_{\mu\nu}=2g_{\mu\nu}.
\end{equation}
We consider a perfect fluid as matter field,
the stress-energy tensor of which is given by
\begin{equation}
T^{\mu\nu}=(\rho+P)u^{\mu}u^{\nu}+Pg^{\mu\nu}.
\end{equation}
In a spherically symmetric spacetime 
with self-similarity,
where the homothetic vector is neither
parallel nor orthogonal to the four velocity 
of the fluid element,
the line element is written in terms of 
the comoving coordinate $r$ as (Cahill and Taub 1971)
\begin{equation} 
ds^{2}=-e^{\sigma(\xi)}dt^{2}
+e^{\omega(\xi)}dr^{2}
+r^{2}S^{2}(\xi)(d\theta^{2}
+\sin^{2}\theta d\phi^{2}),
\end{equation}
where $\xi=t/r$.
In this coordinate system, the dimensionless quantity $h$
such as $e^{\sigma}$, $e^{\omega}$ and $S$ has scale-invariance 
in the following sense:
\begin{equation}
h(t,r)=h(at,ar)=h(\xi),
\end{equation}
where $a> 0$ is an arbitrary constant.
For a self-similar spacetime, the density field $\rho$
is given via the Einstein equations by
\begin{equation}
\rho(t,r)=a^{2}\rho(at,ar).
\end{equation}
It implies that the density field is written as
\begin{equation}
\rho(t,r)=\frac{1}{8\pi r^{2}}\eta (\xi).
\end{equation}
Therefore, the time evolution is equivalent with 
the scale transformation. 
An example of the time evolution of the density field is displayed in 
Fig.~\ref{fg:similarity}.

The above definition of self-similarity can be 
generalized. One possibility is to discretize the above definition:
the dimensionless quantity $h$ satisfies
\begin{equation}
h(t,r)=h(at,ar),\quad a=e^{n\Delta}, 
\end{equation}
where $n=0, \pm 1,\pm 2, \cdots$.
This kind of self-similarity is referred to as 
{\em discrete self-similarity}.
The discrete self-similar solution has been found 
to be important in the critical behavior 
observed in the gravitational collapse of 
a massless scalar field (Choptuik 1993)
and gravitational waves (Abrahams and Evans 1993).
In this context, the originally defined self-similarity is referred to as
{\em continuous self-similarity}.

Another possible generalization is to allow 
the dimensionless quantity $h$ to be a function
of $r/t^{\alpha}$.
The generalization along this direction was 
done by Carter and Henriksen (1989,1991)
and is referred to as {\em kinematic self-similarity}.
The kinematic self-similarity is covariantly 
defined by the existence 
of kinematic self-similarity vector, which satisfies
\begin{equation}
{\cal L}_{\xi}g_{\mu\nu}=2\delta g_{\mu\nu}, \quad 
{\cal L}_{\xi}u_{\mu}=\alpha u_{\mu},
\end{equation}
where $\delta$ and $\alpha$ are arbitrary constants.
See Coley (1997), Benoit and Coley (1998)
and Maeda et al. (2002a, 2002b, 2003) for more details
of kinematic self-similarity and kinematic self-similar solutions.
In this context, the originally defined self-similarity is referred to as
{\em first-kind self-similarity}.

\section{Self-similar solutions}
\label{sec:selfsimilarsolutions}
Hereafter we mainly concentrate on the originally defined
self-similarity 
and refer to it simply as self-similarity unless
it is confusing.
For a perfect fluid, the Einstein equation requires that 
the equation of state be given by the following form:
\begin{equation}
P=k\rho,
\end{equation}
where $k$ is constant (Cahill and Taub 1971).
The fluid is referred to as a dust fluid for $k=0$,
a radiation fluid for $k=1/3$
and a stiff matter for $k=1$.
There are infinitely many self-similar solutions. 
These solutions are obtained by solving ODEs.
The classification of self-similar solutions 
of this system has been done 
in the homothetic approach
by Goliath, Nilsen and Uggla (1998a, 1998b)  
and in the comoving approach
by Carr and Coley (2000a, 2000b).
For the complementarity of these two approaches,
see Carr et al. (2001).
For a brief review of self-similar solutions and, in particular, 
the classifications of them, see Carr and Coley (1999). 

In obtaining physical solutions, it will be required that 
the solution be obtained from regular initial conditions.
In particular, the physical solution must have 
the initial data with regular center.
For $0<k<1$, the central density parameter $D_{0}$,
which is defined by
\begin{equation}
D_{0}\equiv 4\pi \rho(t,0)t^{2},
\end{equation}
well parametrizes the regular center of self-similar solutions, 
where $t$ is chosen to
be negative. 
For the regular center of self-similar solutions,
the $D_{0}$ is finite and does not depend on $t$ because
\begin{equation}
D_{0}=\lim_{r\to 0+}4\pi \rho(t,r)t^{2}=\lim_{\xi\to -\infty}\frac{1}{2}
\xi^{2}\eta(\xi).
\end{equation}  
For $0<k<1$, the ODEs have a singular point which 
is called a sonic point, at which the relative speed $|V|$ 
of the fluid element to the curve $\xi=\mbox{const}$ 
is equal to the sound speed $\sqrt{k}$.
The matching condition beyond the sonic point results in nontrivial 
reduction of self-similar solutions.
The reduction of the self-similar solutions was
demonstrated by Ori and Piran (1990).
If we require regularity, by which we mean the continuity
of first derivatives, 
the possible solutions make a band structure 
in the space of all solutions.
The $D_{0}$ axis is divided into allowed ranges
and forbidden ranges (for which the solution terminates
at the sonic point).
If we further require analyticity (Taylor-series expandability), 
the possible solutions are discretized.
The regular center is also analytic, where
it should be noted that 
the analyticity at the center is defined with respect to the local 
Cartesian coordinates. 
However, only a discrete set of values for $D_{0}$ is allowed 
for the analyticity of the sonic point.

This complicated structure of physical solutions is due to
the character of the sonic point as a critical point of the autonomous
system, which was studied by Bicknell and Henriksen (1978).
The sonic point regarded as a critical point of the 
autonomous system of ODEs after an appropriate 
transformation and is considered in the two-dimensional 
subspace of the whole state space.
As a result, sonic points are classified into 
nondegenerate nodes, degenerate nodes, saddles and foci,
among which the foci are unphysical.
For every physical sonic point, there are two 
possible directions along which solutions can 
cross the sonic point, and there are two 
analytic solutions corresponding to these 
two directions.
For a nondegenerate node, the two directions
are called primary and secondary ones.
For the former, there are one-parameter family of 
regular solutions including one of the analytic solutions,
while for the latter, another analytic 
solution alone is possible.
For a degenerate node, the two directions are 
degenerate and there are one-parameter family of 
regular solutions including the one analytic solution.
For a saddle one, all that are regular at the 
sonic point are the two analytic solutions alone.

\begin{figure}[htbp]
\begin{center}
\includegraphics[height=7cm,width=7cm]{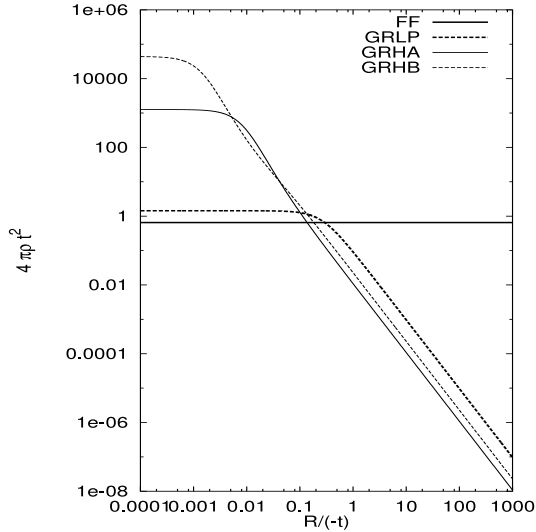}
\caption{\label{fg:DR}
Density profiles for self-similar solutions with analyticity for $k=0.01$.
`FF', `GRLP' and `GRHA'
denote the flat Friedmann solution, general relativistic 
Larson-Penston solution or Ori-Piran solution and 
general relativistic Hunter (a) solution
or Evans-Coleman solution, respectively.} 
\end{center}
\end{figure}

The numerical solutions with  
analytic initial data were 
investigated by Ori and Piran (1987, 1990) and Foglizzo and 
Henriksen (1993).
This discrete set of solutions includes 
the flat Friedmann solution, general relativistic 
counterpart of the Larson-Penston solution,
and that of the Hunter family of solutions.
However, it is noted that the whole set of self-similar solutions
with analytic initial data is still uncertain
because of the numerical difficulty.
Hereafter we refer to the general relativistic Larson-Penston solution
and Hunter (a) solution as the Ori-Piran solution and Evans-Coleman 
solution for convenience.
The most researched nontrivial self-similar solutions are
the Ori-Piran and Evans-Coleman solutions. 
For the character of the sonic point,
the Ori-Piran solution crosses a degenerate 
node along the secondary direction
for $0<k\alt 0.036$, 
a degenerate node for $k\simeq 0.036$
and a nondegenerate node along the primary direction
for $0.036\alt k<1/3$ (Ori and Piran 1990, Harada 2001).
Carr et al. (2001) investigated the detailed nature 
of the Evans-Coleman solution for $0<k<1$.  
The Evans-Coleman solution exists for $0<k\le 1$
(Brady et al. for $k=1$).
It crosses a saddle for $0<k\alt 0.41$,
a nondegenerate node along the secondary direction for $0.41 \alt k \alt
0.89$, a degenerate node for $k\simeq 0.89$
and a nondegenerate node along the primary direction for $0.89 \alt k<1$.
The Ori-Piran solution describes coherent collapse, while
the Hunter family of solutions some oscillative collapse.
The Hunter family of solutions are well labelled by 
the number of zeroes in their velocity fields.
Ori and Piran (1987, 1990) also pointed out that a naked singularity 
forms in the Ori-Piran solution for $0<k \alt  0.0105$,
which may have some implication to 
cosmic censorship proposed by Penrose (1969, 1979).
(See Harada, Iguchi and Nakao (2002) for recent works 
in naked singularity physics.)
Actually, the general relativistic counterpart of 
the Hunter family of solutions with more than one oscillation
can also have a naked singularity (Ori and Piran 1990, 
Foglizzo and Henriksen 1993). 
It is not clear whether nontrivial solutions other than
the Evans-Coleman solution exist
for every value of $k \in (0,1]$, 
although numerical studies suggest that
the structure of the solutions is common 
at least for $0<k\le 9/16$ (Ori and Piran 1990, 
Foglizzo and Henriksen 1993).
The density profiles for these solutions are shown in Fig.~\ref{fg:DR}.

For $k=0$, i.e., a dust fluid, things are somewhat different.
Ori and Piran (1990) studied self-similar dust solutions
with regular center.
There is one-parameter family of solutions.
The regularity at the center requires that 
the central density parameter $D_{0}$ be $2/3$.
The solutions are parametrized not by the central density parameter
but by the central third-order derivative of the density field.
However, this regular center is not analytic except for 
the flat Friedmann solution because 
the third-order derivative of the density field is not zero.
Therefore, if we require that a physical solution has
analytic initial data, the only allowed solution 
is the flat Friedmann solution. 
There is no sonic point for the dust case.
See Carr (2000) for more general class of 
self-similar dust solutions.

\section{Critical behavior}
\label{sec:criticalbehavior}
\begin{figure}[htbp]
\begin{center}
\begin{tabular}{cc}
(a)\includegraphics[height=7cm,width=7cm]{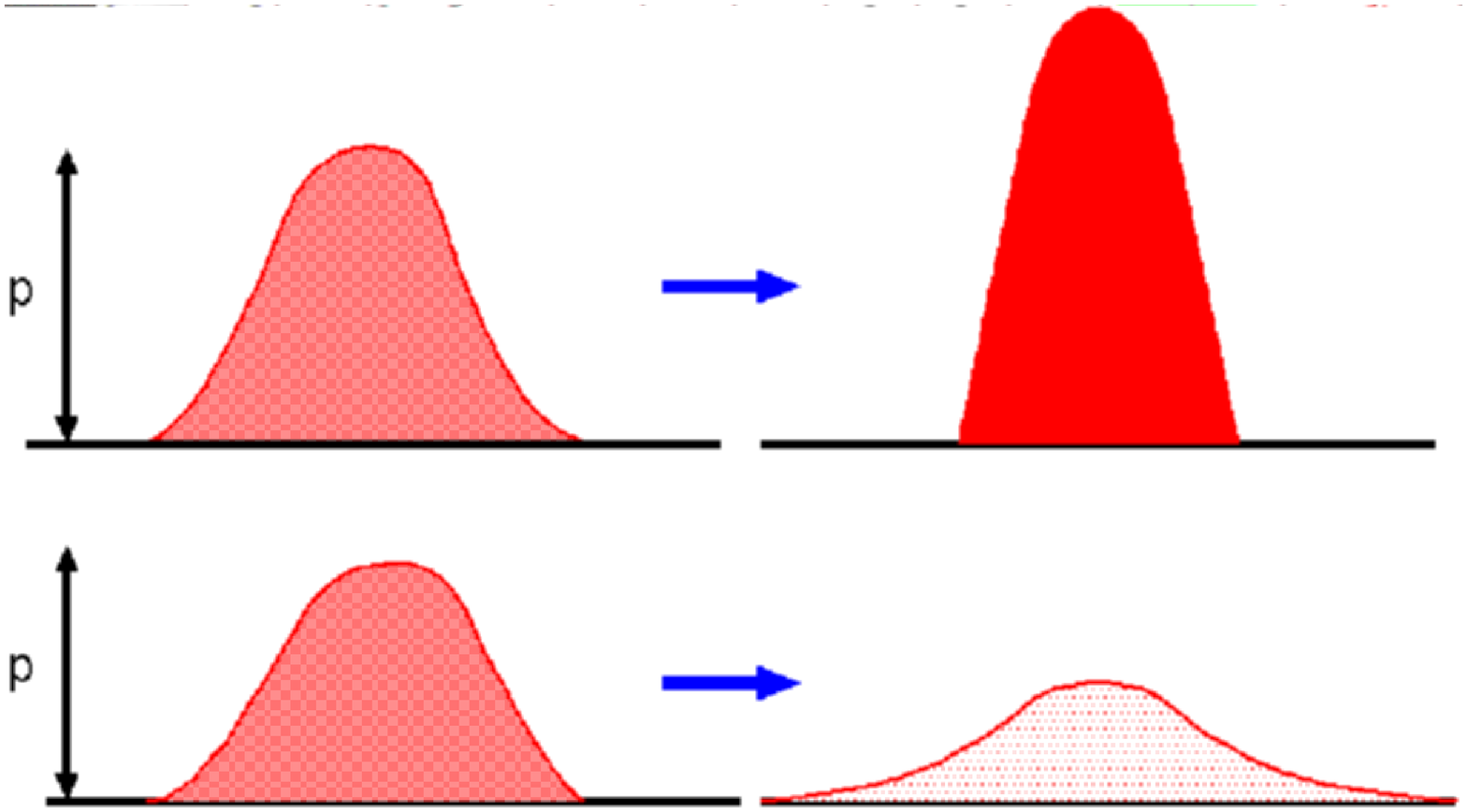} &
(b)\includegraphics[height=7cm,width=7cm]{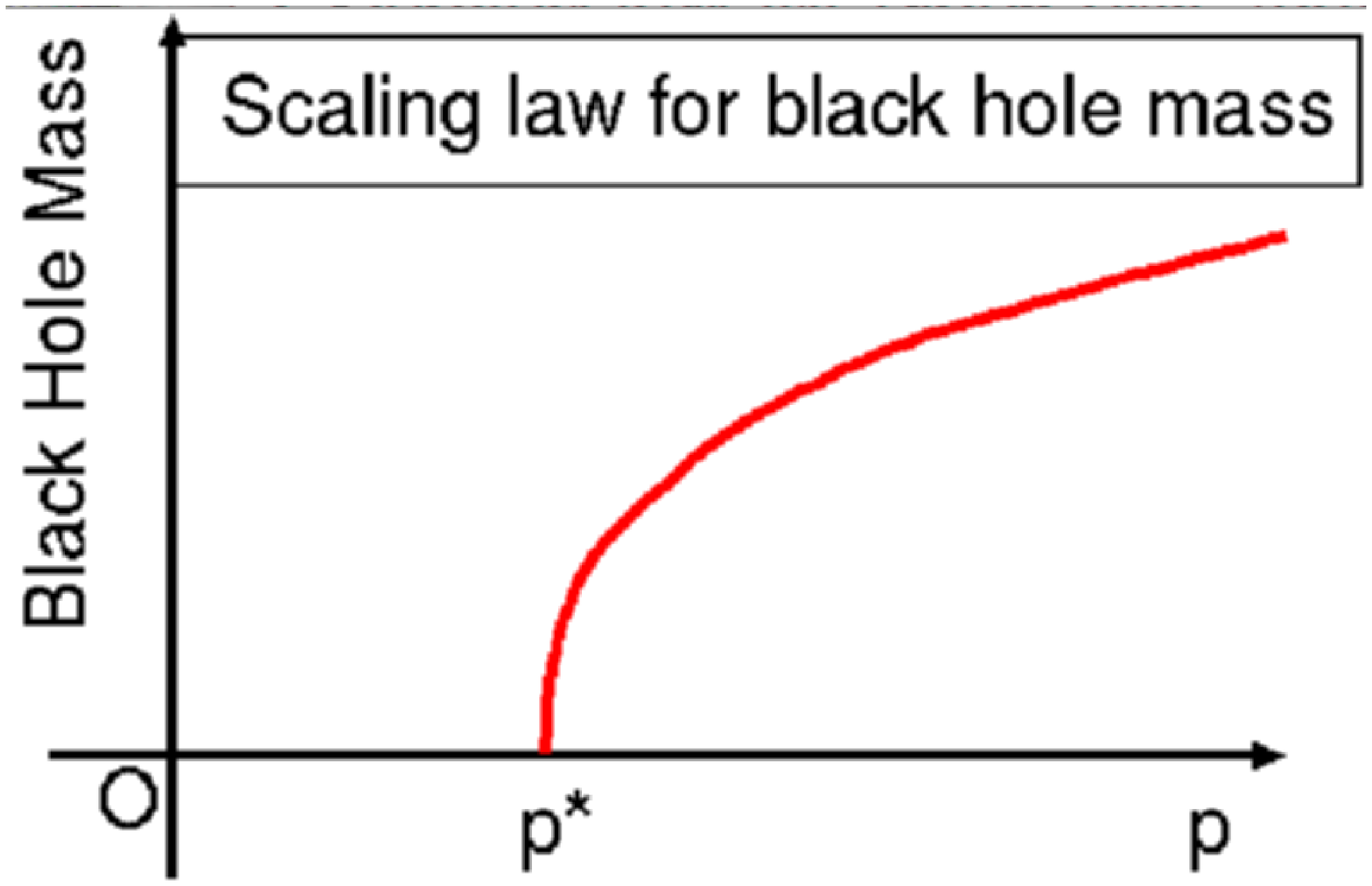} \\
\end{tabular}
\caption{\label{fg:criticalbehavior}
Schematic pictures for the 
critical behavior in gravitational collapse.
(a) indicates that the gravitational collapse has 
critical nature in general. In (b), the scaling law for the formed 
black hole is schematically displayed.}
\end{center}
\end{figure}

The critical behavior in gravitational collapse 
was numerically discovered by Choptuik (1993)
in the spherical system of a massless scalar field.
For a one-parameter family of initial data sets for this system, 
a critical value is found beyond which 
the evolution results in black hole formation 
(see Fig.~\ref{fg:criticalbehavior}(a)).
For near-critical values, the critical behavior
is observed.
The solution first tends to a discrete self-similar solution, which 
is called a critical solution, and diverges away after that.
For near-critical and supercritical collapse, the mass 
of the formed black hole follows a power law as
\begin{equation} 
M_{BH}\propto |p-p^{*}|^{\gamma},
\end{equation}
where the constant $\gamma$ is called a critical exponent
(see Fig.~\ref{fg:criticalbehavior}(b)). 
The critical exponent for this system was estimated to be 
$\approx 0.37$.
After this discovery, a similar behavior was observed
in various systems, such as gravitational waves
(Abrahams and Evans 1993), a radiation fluid (Evans and Coleman 1994), 
a perfect fluid, a stiff matter (Neilsen and Choptuik 2000) and so on.
Then, it has been found that the critical solution 
can be a discrete self-similar solution
or continuous self-similar solution, depending on the system.
For the fluid system including the radiation fluid system, 
the critical solution has continuous self-similarity.
For the system of a radiation fluid, 
Evans and Coleman (1994) estimated 
the critical exponent
to be $\approx 0.36$ and also found that the critical solution 
obtained as a result of numerical simulations of the PDEs
agrees very well with one of the continuous self-similar solutions
which are obtained by integrating the ODEs. 
(That is why we refer to this self-similar solution as
the Evans-Coleman solution in this article.
We shall avoid the term `critical solution' to refer to this solution 
because whether this solution is a critical solution 
may depend on the system, for example, whether we allow nonspherical
perturbations or not, as we will mention later.)
It has been subsequently found 
that the value of the critical exponent is not universal but 
depends on the parameter $k$
which parametrizes the equation of state $P=k\rho$.  

Koike, Hara and Adachi (1995) gave a clear picture 
for the critical behavior, which is called a
renormalization group approach.
According to this picture, the critical solution 
is assumed to be a self-similar solution with analytic initial data
with a single unstable mode.
We consider the time evolution of the solution with respect to
the self-similarity coordinates $\tau\equiv -\ln |t|$ and 
$x\equiv -\ln |\xi|$.
Then, the time evolution is regarded as a flow
in the space of functions of $x$. 
The self-similar solution with a single unstable mode,
which is a fixed point in the space of functions of $x$,
has a stable manifold of codimension one.
This implies that a one-parameter family of initial data
generically has intersection with the stable manifold.
The collapse which develops from the 
initial data set which is the intersection corresponds to 
the critical collapse, while near-critical collapse develops
from the initial data set near the intersection.
After a long time, the deviation of the near critical collapse
from the critical solution
is dominated by the single unstable mode, i.e., 
\begin{equation}
h(\tau,x)\approx H_{ss}(x)+(p-p^{*}) e^{\kappa \tau}F_{rel}(x),
\end{equation}
where $H_{ss}$ is the critical solution and $\kappa$ and $F_{rel}$
are the eigenvalue and eigenfunction of the single unstable mode.
Since the apparent horizon appears when the deviation becomes $O(1)$,
the mass of the black hole is estimated as
\begin{equation}
M=O(r)=O(e^{-\tau})\propto |p-p^{*}|^{\gamma}
\end{equation}
where $\gamma=\kappa^{-1}$.
The eigenvalue and eigenfunction of the unstable mode 
are determined by the linearized Einstein equation with 
the regularity condition both at the center and 
at the sonic point.
Koike, Hara and Adachi calculated the critical exponent
to be $\approx 0.3558019$ by the eigenvalue analysis,
which agrees very well with the estimated value $\approx 0.36$
from the numerical simulation. 

Critical phenomena in gravitational collapse have been 
observed and widely researched in a great variety of systems. 
The implication to the formation of 
primordial black holes was also discussed.

One of the interesting systems is the spherical system 
of a stiff matter.
Neilsen and Choptuik (2000) observed 
the critical behavior in the stiff-matter system
and found that the critical solution is 
a continuous self-similar solution.
They also estimated the critical exponent to be $0.96\pm 0.02$.
Brady et al. (2002) confirmed that the Evans-Coleman 
solution for $k=1$ has a single unstable mode and 
that the critical exponent derived from the eigenvalue
analysis agrees well with that estimated from the result
of numerical simulations of the PDEs.
This critical behavior for the stiff-matter system 
seems to be totally different from that for the 
scalar-field system. 
On the other hand, a massless scalar field is equivalent 
with a stiff matter under certain circumstances 
(Madsen 1988, Brady et al. 2002).
Brady et al. (2002) pointed out that the critical solution,
which is discrete self-similar,
for the system of a massless scalar field
cannot be constructed from the stiff matter
because the equivalence is broken.

See Gundlach (1998, 1999) for a recent review of 
critical phenomena in gravitational collapse.
\section{Stability}
\label{sec:stability}
The stability of self-similar solutions against 
spherically symmetric perturbations has called much attention
in the context of the critical behavior in gravitational collapse
since Koike, Hara and Adachi (1995) showed that the critical behavior
is understood in terms of the behavior of 
the unstable mode of the critical solution.
 
\begin{figure}[htbp]
\begin{center}
\begin{tabular}{cc}
\includegraphics[width=7cm,height=7cm]{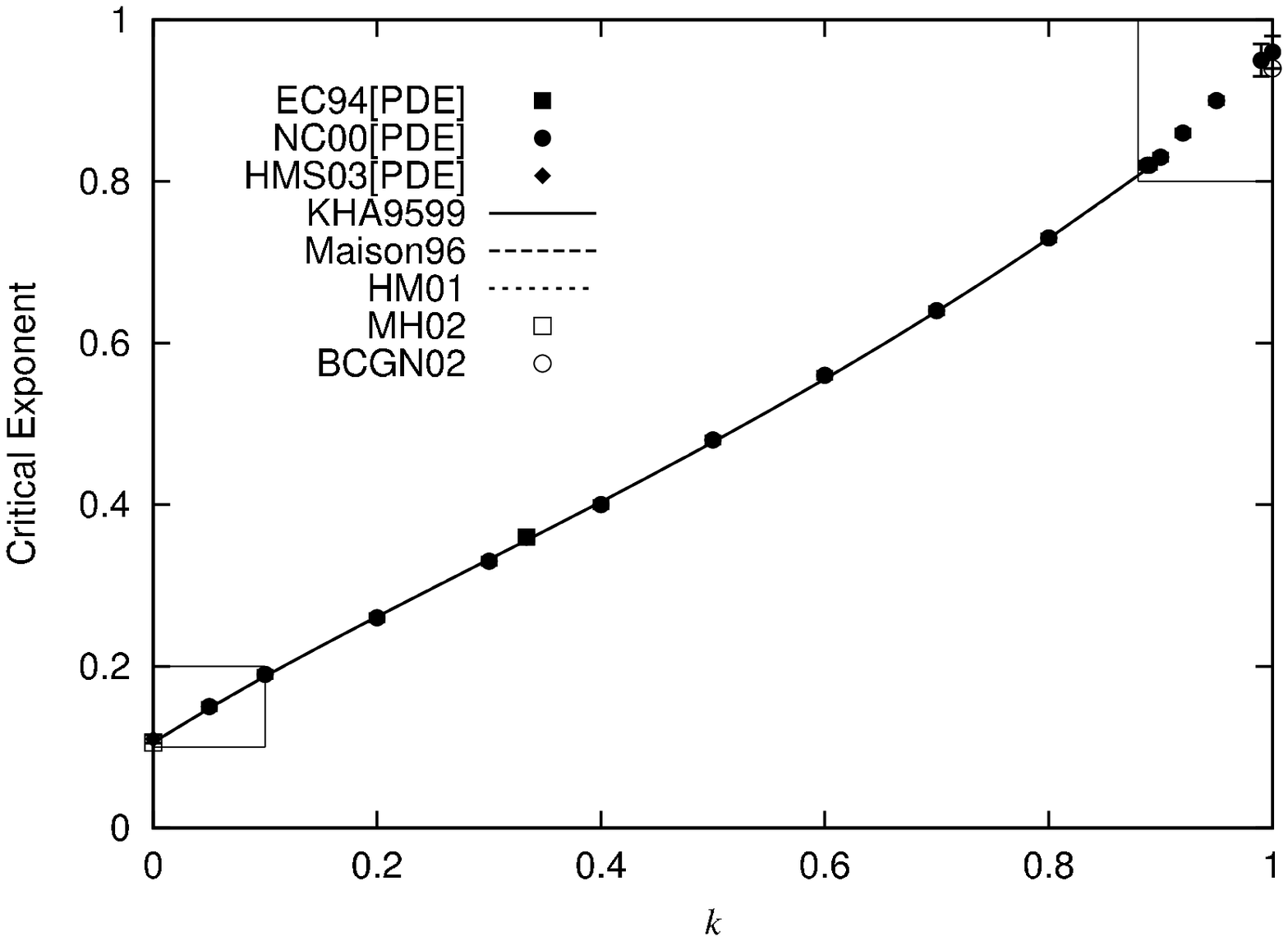} &
\includegraphics[width=7cm,height=7cm]{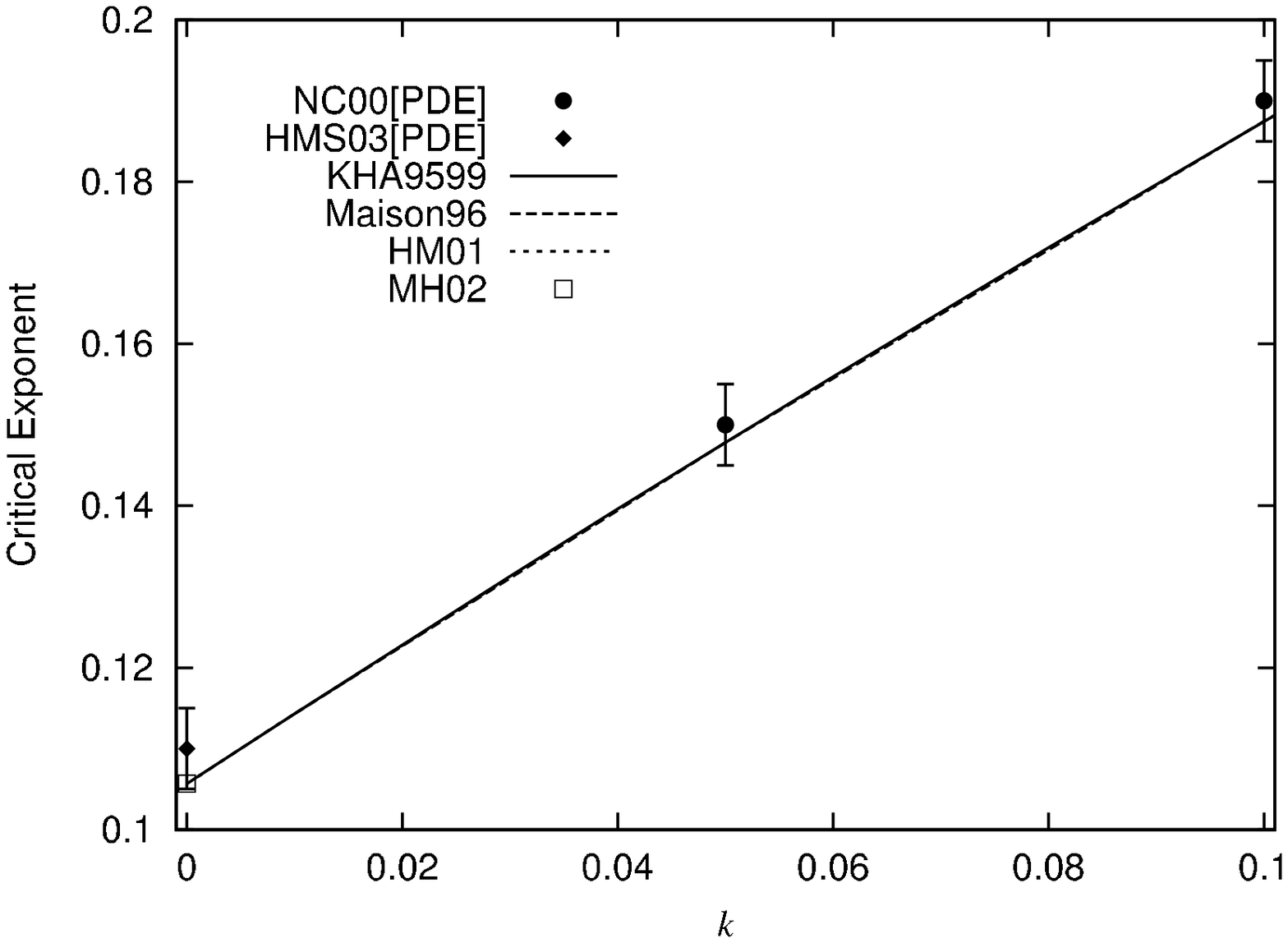} \\
\includegraphics[width=7cm,height=7cm]{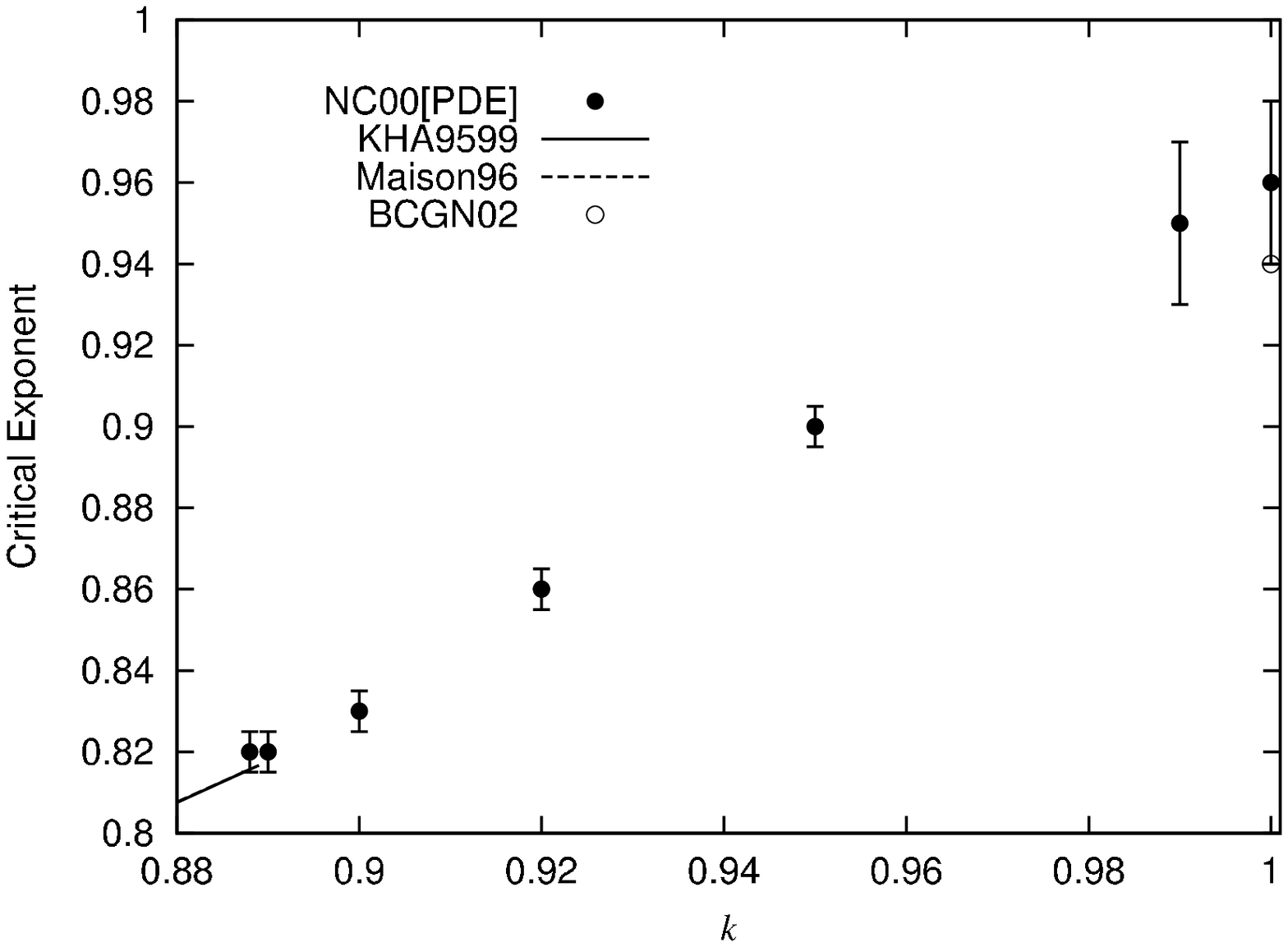} & \\
\end{tabular}
\end{center}
\caption{\label{fg:criticalexponent}
Critical exponents for various value of $k$. 
The plots with the label ``[PDE]'' were obtained as a result of 
numerical simulations of PDEs. All other plots were
obtained by the eigenvalue analyses. The references are
the following: EC94 = Evans and Coleman (1994), 
NC00 = Neilsen and Choptuik (2000), 
HMS03 = Harada, Maeda and Semelin (2003), 
KHA9599 = Koike, Hara and Adachi (1995, 1999),
Maison96 = Maison (1996), 
HM01 = Harada and Maeda (2001),
MH02 = Maeda and Harada (2002), 
BCGN02 = Brady et al. (2002).
The error bars attached with the plots obtained from 
numerical simulations are determined from the significant digits 
when the numerical errors are not explicitly given 
in the reference.
For KHA9599, Maison96 and HM01, the plots are replaced by curves
because there are many plots and numerical errors are
considered to be very small. Actually, these curves are hard 
to discriminate from each other. 
The plots for Newtonian case are plotted at $k=0$
because the Newtonian system is regarded as the $k\to 0$ 
limit of general relativity.
}
\end{figure}

There are two types of spherically symmetric 
perturbations, analytic modes and kink modes.
First we consider the analytic modes.
The perturbation for a dimensionless quantity $h$ is written as
\begin{equation}
h(\tau,x)=h_{0}(x)+\delta h(\tau,x).
\end{equation}
After the linearization with respect to the perturbation, 
the perturbation $\delta h$ is separated in two factors as
\begin{equation}
\delta h(\tau,x)=\epsilon e^{\kappa \tau}F(x).
\end{equation}
For $0<k\le 1$, when
the regularity condition is imposed 
both at the center and at the sonic point,
the $\kappa$ must take a discrete set of values.
In other words, the allowed values for $\kappa $ are determined 
as eigenvalues.
If $\kappa>0$, the mode is growing and  
referred to as an unstable mode.
The stability of the Evans-Coleman solution 
has been investigated by Koike, Hara and Adachi (1995) for $k=1/3$ and 
by Maison (1996), Koike, Hara and Adachi (1999), 
Harada and Maeda (2001) for more general value 
of $k\in (0,1)$, and Brady et al. (2002) 
for $k= 1$.
All these works have shown that the Evans-Coleman solution 
has a single analytic unstable mode for $0<k\le 1$.
For the Evans-Coleman solution, which has a single unstable mode,
the inverse of the eigenvalue of the unstable mode is interpreted as
the value of the critical exponent, 
as we have seen in Sec.~\ref{sec:criticalbehavior}.
The results are summarized in Fig.~\ref{fg:criticalexponent}.
The critical exponents directly estimated from the mass-scaling law 
obtained after the numerical simulations of the PDEs are also plotted.

The stability of other self-similar solutions with 
analytic initial data were
investigated by Koike, Hara and Adachi (1999) and 
Harada and Maeda (2001).
Then, it has been found that 
the Ori-Piran solution have no unstable mode, while 
the general relativistic counterpart of the Hunter-family solutions 
have unstable modes, the number 
of which agrees with the number of zeroes in the 
velocity field at least for the solutions examined there.
Since the nonuniqueness of unstable modes implies that 
the codimension of the stable manifold is greater than one,
it is expected that
the approach to the self-similar solution is realized
only as a result of fine-tuning more than one parameters
simultaneously.
This is the reason why the numerical simulations of the PDEs
do not hit the self-similar solution with more than one
unstable modes.
On the other hand, since the Ori-Piran solution has no 
unstable mode, it is expected that generic evolution go towards this 
solution without any fine-tuning. 
In the next section, we will see that actually this is the case. 

\begin{figure}[htbp]
\begin{center}
\begin{tabular}{cc}
\includegraphics[width=7cm,height=7cm]{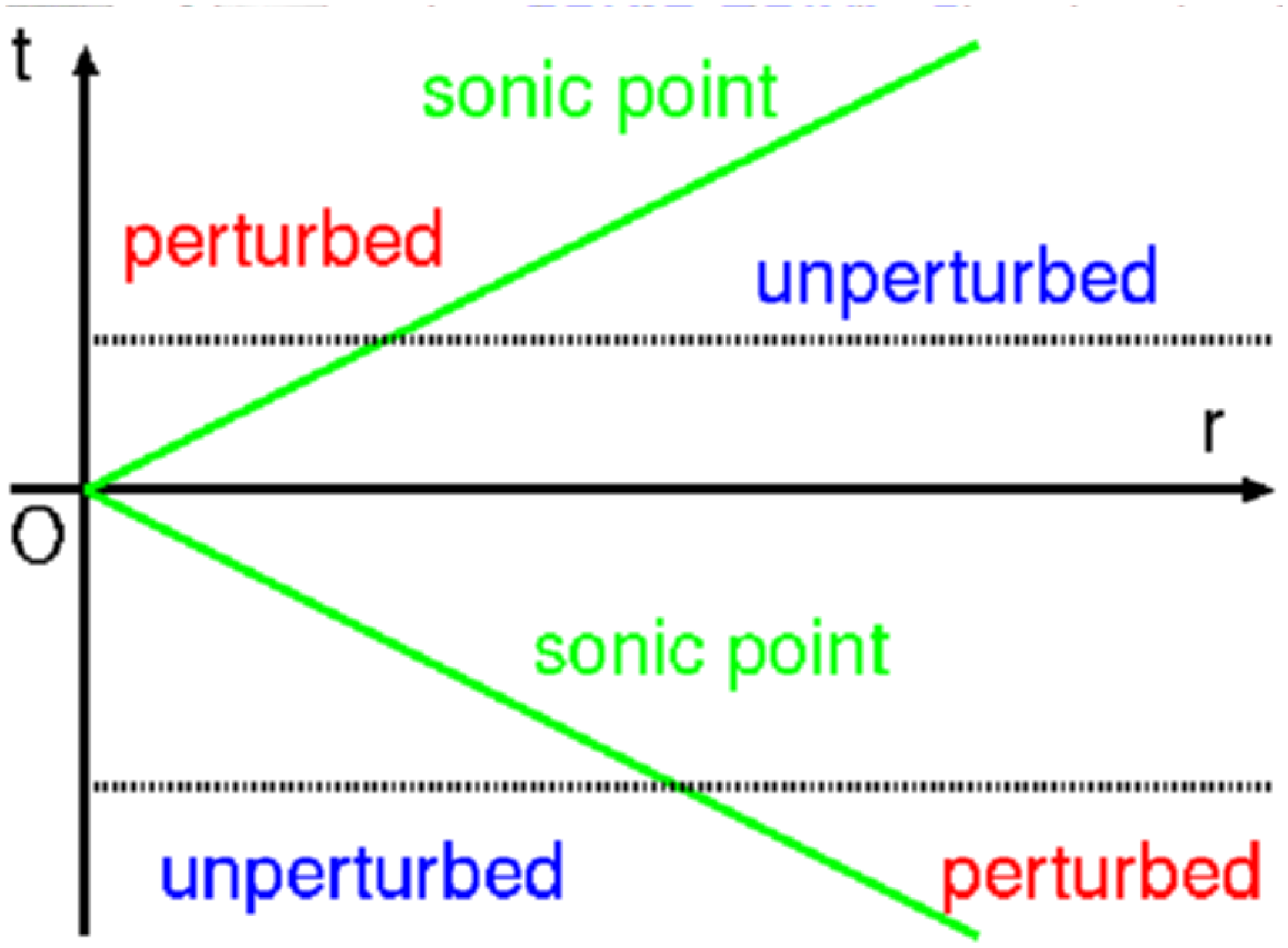} &
\includegraphics[width=7cm,height=7cm]{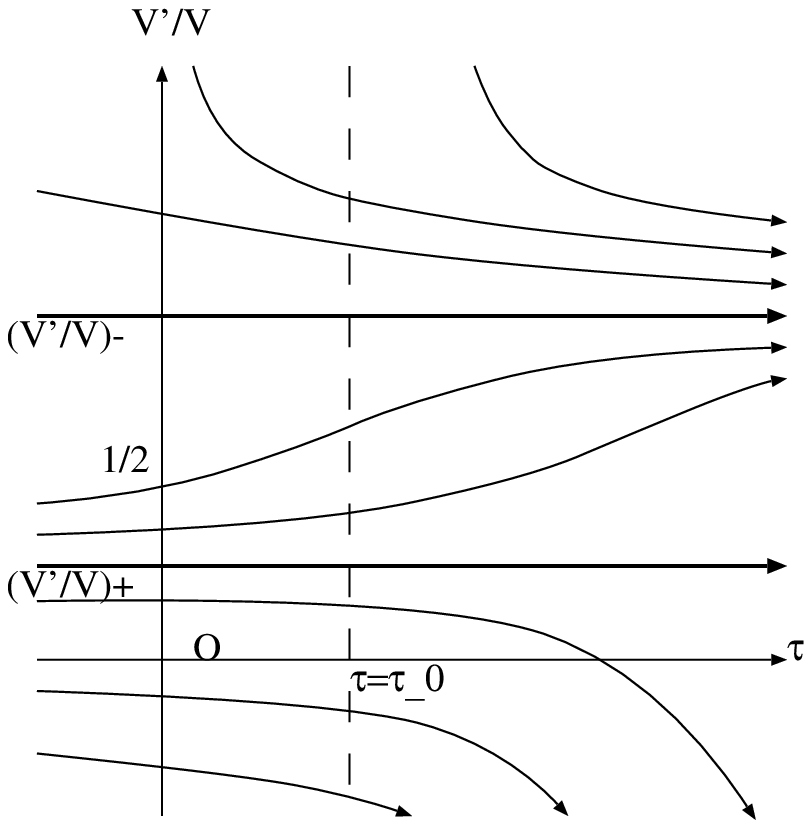} \\
\end{tabular}
\end{center}
\caption{\label{fg:kink}
(a) Initial setting of kink-mode perturbations.
For $t<0$, the initial perturbation is put 
only outside the sonic point.
(b) The nonlinear evolution of the perturbation
is depicted. $+$ and $-$ denote the primary and secondary directions,
respectively. The density gradient is proportional 
to the velocity gradient $V'/V$, where the prime denotes the 
partial derivative with respect to $x=-\ln |\xi|$. 
See Harada (2001) for details.}
\end{figure}

Next, for the kink modes, the existence in general relativity was
expected by Ori and Piran (1990) and demonstrated by Harada (2001).
For the analysis, we consider the initial perturbations which 
satisfy the following conditions: (For a while, we assume
$t<0$ and later will mention the case for $t>0$.)
(1) The initial perturbations vanish inside the sonic point.
(2) The density field is continuous.
(3) The density gradient field is discontinuous at the 
sonic point, although it has definite one-sided values 
inside and outside.
See Fig.~\ref{fg:kink}(a) for the initial setting of 
perturbations.
The evolution of the density gradient at the sonic point
turns out to be local, which simplifies the analysis.
The analysis of the kink mode is possible in full order.
Then, it is found that the stability against the kink mode
is closely related to the classification of the sonic point,
where the instability is defined by the blow up of the 
density gradient in a finite interval in terms of 
the time coordinate $\tau$.

The result is as follows. All primary-direction nodal-point solutions,
degenerate-nodal-point solutions, and saddle-point anti-transsonic 
solutions are unstable against the kink mode,
while all secondary-direction nodal-point solutions, 
and saddle-point transsonic solutions are stable against
the kink mode.
The situation is reversed for $t>0$ except for the case
of degenerate-nodal-point solutions, which are unstable 
against the kink mode both for $t>0$ and $t<0$.
Figure~\ref{fg:kink}(b) shows the nonlinear evolution
of the kink mode for a sonic point which is 
a nondegenerate node, as a typical example.

When we apply this criterion to the known solutions, 
we obtain the following results for the stability against
the kink mode.
The flat Friedmann (collapsing) solution is
unstable for $0<k<1/3$ while stable 
for $1/3\le k<1$.
The Ori-Piran solution is stable for $0<k\alt 0.036$
while unstable for $0.036\alt k<1/3$.
The Evans-Coleman solution is stable for $0<k\alt 0.89$
while unstable for $0.89\alt k<1$.
Moreover, all regular but nonanalytic (collapsing) solutions 
are unstable.

The kink instability of the Evans-Coleman solution 
for $0.89\alt k<1$ will affect the critical behavior 
of a perfect fluid. 
However, Neilsen and Choptuik (2000) observed the 
critical behavior even for $0.89\alt k\le 1$
which is considered to be continuous
extrapolation of that for $0<k\alt 0.89$.
It has not been explicitly shown but is suggested that 
the Evans-Coleman solution has a single analytic 
unstable mode even for $0.89\alt k<1$ from their result.
Moreover, it was also reported that the Evans-Coleman solution 
has a single analytic unstable mode for $k=1$ (Brady et al. 2002).
About the reason why the kink instability did not
affect the PDE calculations by Neilsen and Choptuik (2000),
Harada (2001) speculated several possibilities.

The stability of the Evans-Coleman 
solution against nonspherical perturbations was investigated
by Gundlach (2002). The result is that the Evans-Coleman solution
is stable against nonspherical perturbations for $1/9<k\alt 0.49$
while is unstable for $0<k<1/9$ or $0.49\alt k <1$.
For the stability of other self-similar solutions with analytic 
initial data against nonspherical perturbations, 
little has been known until now.
\section{Attractor}
\label{sec:attractor}
\begin{figure}[htbp]
\begin{center}
\begin{tabular}{cc}
(a)\includegraphics[width=7cm,height=7cm]{d001} &
(b)\includegraphics[width=7cm,height=7cm]{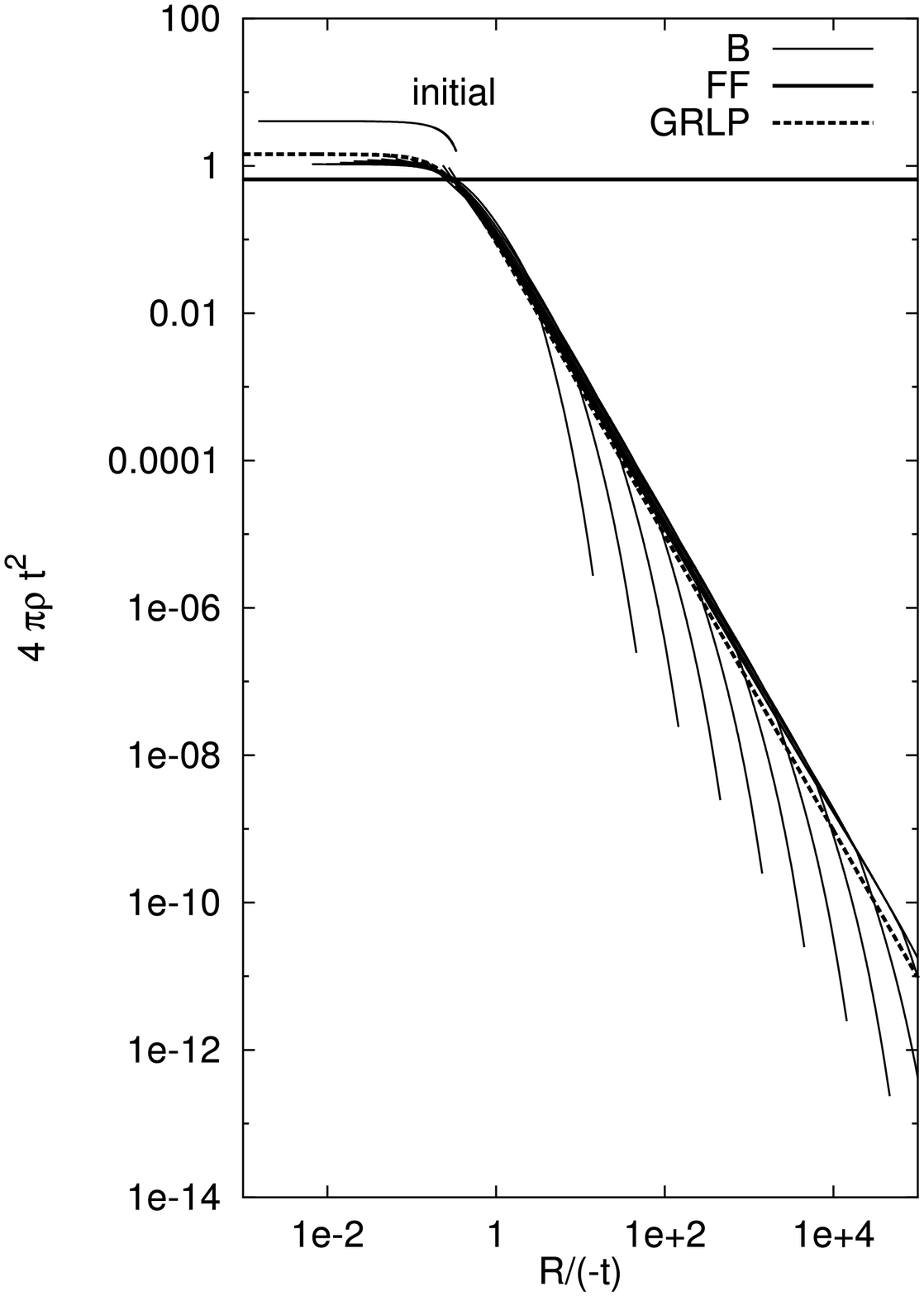} \\
\end{tabular}
\end{center}
\caption{\label{fg:attractor}
In (a), the time evolution of the central density parameter $D_{0}$
is depicted 
as the collapse proceeds and the central density increases.
In (b), the snapshots of density profiles is compared with the density 
profile of self-similar solutions.
These are obtained by the numerical simulations of the PDEs for $k=0.01$.}
\end{figure}

As seen in the previous section, 
we have at least one self-similar solution which has 
no unstable mode.
The solution is the Ori-Piran solution for $0<k\alt 0.036$.
From the discussion based on the renormalization group,
which we have seen in Sec.~\ref{sec:criticalbehavior},
this solution seems to be an attractor which generic
solution approach in the vicinity of the singularity
formation.
The numerical result obtained by Harada and Maeda (2001)
strongly suggests that this is the case.
In their numerical simulation of the PDEs, 
the collapse ended in singularity 
formation at the center for a certain subset 
of initial data sets, where the initial data sets
were prepared without fine-tuning parameters.
Thus obtained numerical solutions were compared
with the self-similar solutions obtained by
integrating the ODEs.
Then, for the numerical solutions 
obtained by numerical simulations of the PDEs,
the good continuous self-similarity is found 
and the profile of the density field agrees very well 
with the Ori-Piran solution, as seen in Fig.~\ref{fg:attractor}.
It was confirmed that
this convergence to the self-similar attractor 
does not depend on the choice of the 
initial density profile.
The above is the results of numerical simulations 
for $0<k\le 0.03$.

However, it is noted that we can consider initial data sets,
from which the solution does not approach the Ori-Piran solution.
For example, it is clear that the 
initial data set for the 
self-similar solutions other than the Ori-Piran
solution does not show the approach to 
the Ori-Piran solution.
Furthermore, if the central region is initially 
homogeneous and sufficiently compact, it can be shown that 
the solution approach the flat Friedmann solution
around the center as the collapse goes towards 
the singularity formation.
It is suggestive that the final fate of 
all these counterexamples 
is also described by self-similar solutions.  
Anyway, the result of numerical simulations strongly suggests
that the set of the counterexamples to the convergence to the attractor 
is zero-measure in the space of initial data sets.
These numerical results are well incorporated in the 
picture of renormalization flow;
the critical solution as a saddle and 
the attractor solution as a sink in the renormalization flow 
of the function space.

This is one realization of the self-similarity hypothesis
in the spherically symmetric gravitational collapse.
Moreover, since a naked singularity develops 
from the analytic initial data in the Ori-Piran solution
for $0<k\alt 0.0105$,
this numerical study strongly suggests that 
the violation of cosmic censorship is generic in spherically 
symmetric spacetime.
It should be noted that
the numerical study based on the null formulation
also obtained qualitatively the same result
(Harada 1998).

From these considerations, it is expected that 
one of self-similar solutions may 
describe the asymptotic behavior of 
generic and/or special solutions in more general situations.
Actually, several other examples are known to exhibit 
that a self-similar solution describes 
the asymptotic behaviors of more general solutions (Carr and Coley 1999).
If the self-similarity hypothesis is true in a large variety of 
systems, not only will
the analysis of the singularity and/or
the asymptotic behavior of the spacetime be much simplified,
but also should this be argued in the context of the 
structure of generic singularity 
(cf, Belinsky, Khalatnikov and Lifshitz 1970). 

\section{Newtonian case}
Newtonian self-similar solutions have been investigated 
mainly in an effort to obtain
realistic solutions of gravitational collapse
leading to the star formation.
Penston (1969) and Larson (1969) 
independently found a self-similar solution,
which describes a gravitationally collapsing isothermal gas
sphere. Here we call this solution the Larson-Penston solution.
Thereafter, Hunter (1977) found a new series of self-similar 
solutions with analytic initial data and 
that the set of such solutions is infinite and discrete.
Since he named the solutions (a), (b), (c), (d) and so on,
here we call these solutions the Hunter (a), (b), (c) and (d)
solutions and so on.
Shu (1977) found other solution and discussed 
its astrophysical application.
Whitworth and Summers (1985) pointed out the existence of 
a new family of self-similar 
solutions with loss of analyticity and 
found that these solutions make a band structure
in the space of all solutions.
When Ori and Piran (1990) extended these Newtonian self-similar solutions
to general relativity, they showed that these Newtonian 
self-similar solutions are obtained as the limit of $k\to 0$
of general relativistic self-similar solutions.
Even for a spherical gas system with the polytropic equation of 
state, it is possible to extend the definition of 
self-similar solutions and 
these solutions were studied by Goldreich and Weber (1980),
Yahil (1983) and Suto and Silk (1988).
However, here we concentrate on the case of isothermal gas.

There are two kinds of perturbation modes, the one is the analytic modes and 
the other is the kink modes, as we have seen in general relativistic
case.
For the former,
the stability of self-similar solutions were studied 
by Hanawa and Nakayama (1997), Hanawa and Matsumoto 
(2000a, 2000b) and Maeda and Harada (2001) 
against spherically symmetric and nonspherical perturbations.
For spherically symmetric perturbations, it is found that 
the homogeneous collapse solution and the Larson-Penston 
solution have no unstable mode, while the Hunter (a), (b), (c), (d)
solutions have one, two, three and four unstable modes, respectively.
It is expected all Hunter-family solutions have more than one 
unstable modes except for the Hunter (a) solution.  

For the kink-mode perturbation, 
Ori and Piran (1988) showed that for $t<0$ 
all primary-direction nodal-point solutions,
degenerate-nodal-point solutions, and saddle-point anti-transsonic 
solutions are unstable against the kink mode,
while all secondary-direction nodal-point solutions, 
and saddle-point transsonic solutions are stable against
the kink mode.
The situation is reversed for $t>0$ except for the case
of degenerate-nodal-point solutions, which are unstable 
against the kink mode both for $t>0$ and $t<0$.
Actually, the above result is completely the same as in 
general relativity.
Applying this criterion against the kink mode, it is found that 
the homogeneous (collapsing) solution is unstable while the Larson-Penston
solution and the Hunter (a) solution are stable.

Since the above analysis indicates that the Hunter (a)
solution has a single unstable mode,
Maeda and Harada (2001) identified that the Hunter (a)
solution is a critical solution in Newtonian collapse. 
The critical exponent calculated from the eigenvalue 
analysis is $\approx 0.10567$, which is appropriate value 
as the $k\to 0$ limit of the critical exponent in general relativity.
Moreover, the character of the Hunter (a) solution is very similar 
to that of the Evans-Coleman solution. For example, for both solutions,
the central region contracts, the surrounding region expanding.
From these observations, Harada and Maeda (2001) 
and Maeda and Harada (2001) identified that 
the Evans-Coleman solution is the general relativistic counterpart 
of the Hunter (a) solution. 
 
Moreover, since the Larson-Penston solution has no unstable 
mode, it is regarded as an attractor solution.
Actually, it has been considered and demonstrated 
that the Larson-Penston solution
describes the generic collapse of an isothermal gas
near the center since the discovery of this solution
(Penston 1969, Larson 1969, Foster and Chevalier 1993, 
Tsuribe and Inutsuka 1999).

\begin{figure}[htbp]
\begin{center}
\begin{tabular}{cc}
\includegraphics[width=7cm,height=7cm]{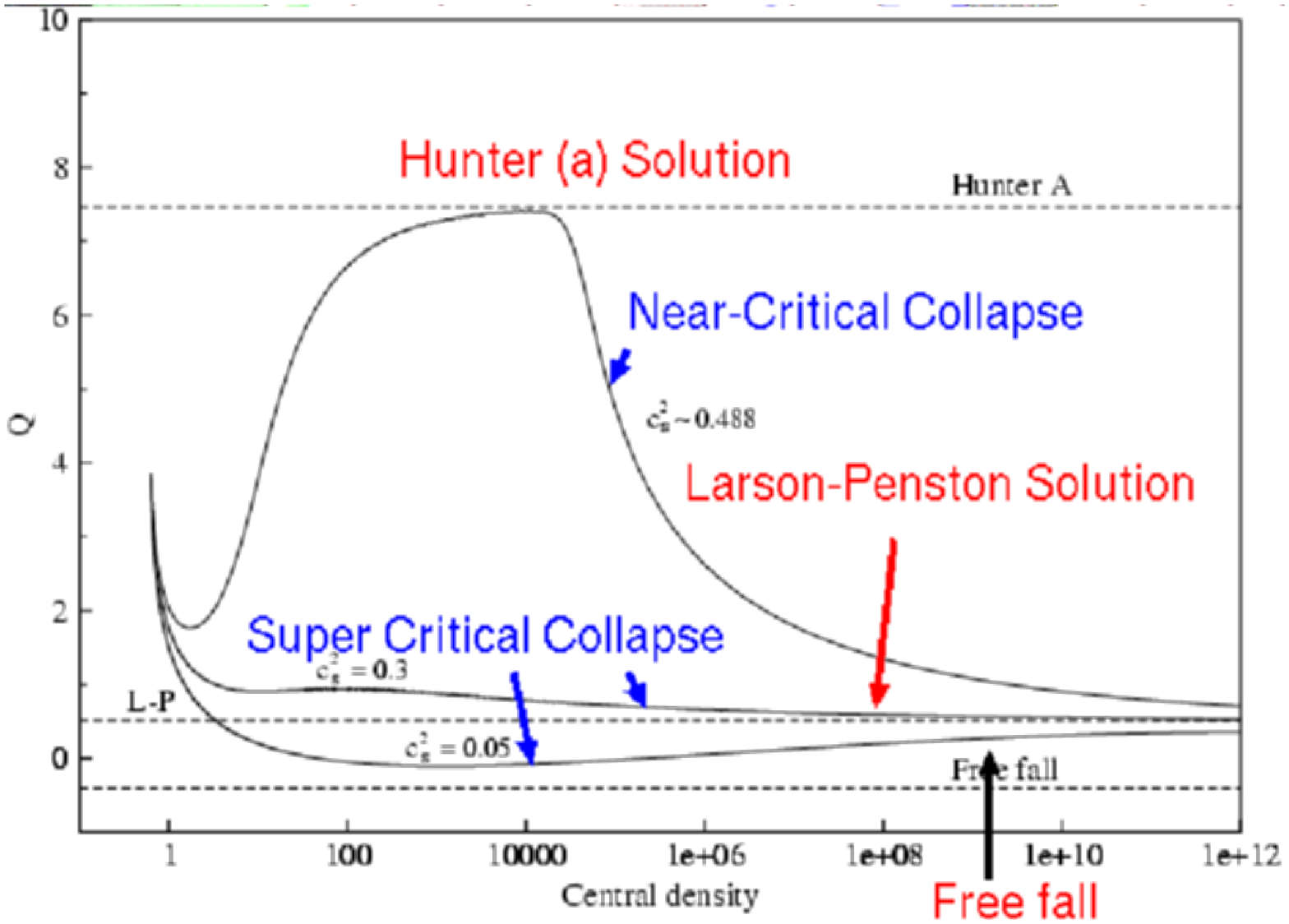} &
\includegraphics[width=7cm,height=7cm]{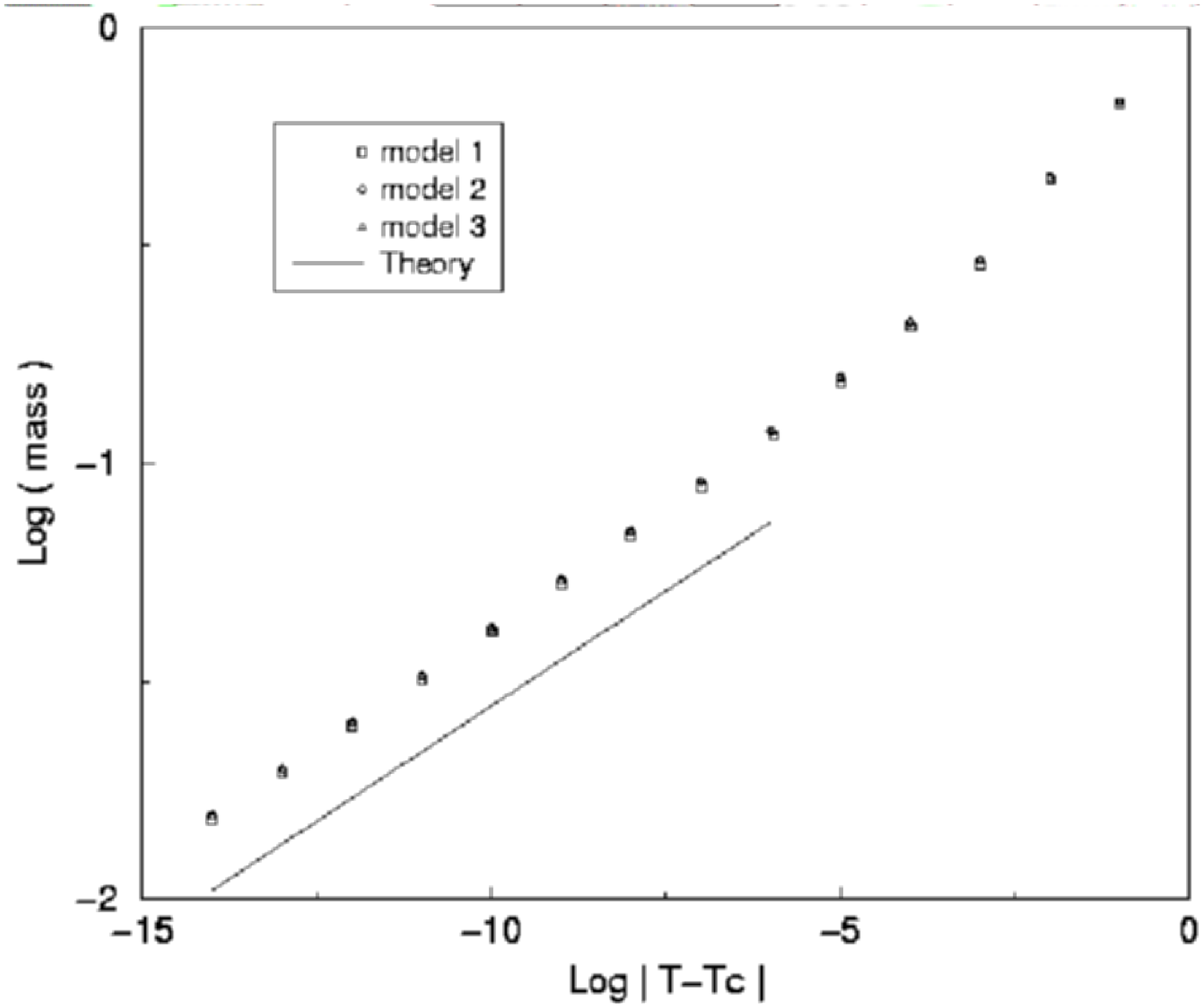}\\
\end{tabular}
\end{center}
\caption{\label{fg:newtoncritical}
In (a), the time evolution of 
the central density parameter $Q=\ln D_{0}=\ln [4\pi \rho(t,0)t^{2}]$
is depicted as a function of the central density.
In (b), the scaling law for the collapsed core is depicted.
In these simulations, the parameter of the one-parameter 
family of initial data sets is chosen to be 
the temperature of the system, the density profile fixed. 
For the details for definitions of quantities in this figure,
see Harada, Maeda and Semelin (2003).}
\end{figure}

To find the critical behavior in Newtonian collapse
and see the relation between the attractor and critical solutions,
Harada, Maeda and Semelin (2003) performed the numerical simulations
of the PDEs which govern the spherical system of an isothermal gas.
The result is as follows:
When a one-parameter of initial data sets is prepared, 
there is a critical value of the parameter.
For the critical value the collapse tends to the Hunter (a)
solution.
For near-critical subcritical case, the collapse first 
approaches the Hunter (a)
solution, thereafter diverges from that and finally disperses away.
For near-critical supercritical case, the collapse first 
approaches the Hunter (a) solution, thereafter diverges from that 
and finally approaches the Larson-Penston solution.
For the solution to approach the Larson-Penston solution,
the parameter fine-tuning is not necessary but the parameter 
only has to be supercritical.
The scaling law appears for the mass of the collapsed core
for the near-critical supercritical case, while
for the maximum density for the near-critical subcritical case.
The critical exponents for these laws were estimated to be 
$\approx 0.11$ and $\approx -0.22$,
which agree very well with the expected values, $\approx 0.10567$ and
$\approx -0.21134$,
respectively.
Figure~\ref{fg:newtoncritical} shows the trajectory of solutions
and the obtained scaling law for the collapsed core mass.

\section{Summary}
\begin{figure}[htbp]
\begin{center}
\includegraphics[width=7cm,height=7cm]{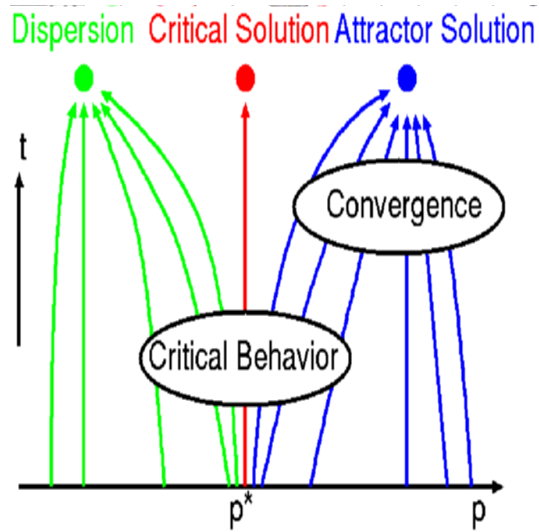}
\caption{\label{fg:fate}
Schematic picture for the fate of gravitational collapse.
This picture applies at least to the spherically symmetric perfect-fluid
system with $0<k\le 0.03$ and to the Newtonian isothermal gas system.}
\end{center}
\end{figure}

The general theory of relativity as well as Newtonian gravity
admits self-similar solutions. 
This is due to the scale-invariance of gravity.
The self-similar solutions are important not only because
they are dynamical and inhomogeneous solutions 
easy to obtain but also 
because they may play important roles 
in the asymptotic behavior of more 
general solutions.
We have seen that the stability analysis of self-similar solutions 
gives a unified picture of the critical behavior and convergence to
attractor in gravitational collapse, which have been both
observed in recent numerical simulations of the PDEs.
Since both the critical behavior and convergence to attractor 
are not only in general relativity but also in Newtonian gravity,
these two are considered to be common characters of gravitational 
collapse physics.
In the above sense, gravitational collapse 
singles out self-similar solutions even if the 
initial condition given is complicated. 
Finally, this article is concluded by 
the schematic picture for gravitational collapse
in Fig.~\ref{fg:fate}.

\begin{thebibliography}{99}
\bibitem{ae1993}
  A.M.~Abrahams and C.R.~Evans,
  Phys. Rev. Lett.
  {\bf 70}, 2980 (1993).
\bibitem{bhl1970}
V.A.~Belinsky, I.M.~Khalatnikov and E.M.~Lifshitz, 
Usp. Fiz. Nauk. {\bf 102}, 463 (1970). 
[English translation in Advances in Physics {\bf 19}, 525 (1970)].
\bibitem{bc1998}
  P.M.~Benoit and A.A.~Coley,
  Class. Quantum Grav.
  {\bf 15}, 2397 (1998).
\bibitem{bh1978b}
  G.V.~Bicknell and R.N.~Henriksen,
  Astrophys. J. {\bf 225}, 237 (1978).
\bibitem{bcgn2002}
P.R.~Brady, M.W.~Choptuik, C.~Gundlach and D.W.~Neilsen,
Class. Quantum Grav. {\bf 19}, 6359 (2002).
\bibitem{CT1971}
  M.E.~Cahill and A.H.~Taub,
  Commun. Math. Phys.
  {\bf 21}, 1 (1971).
\bibitem{carr1993}
  B.J.~Carr, unpublished (1993).
\bibitem{carr2000}
B.J.~Carr,
Phys. Rev. D
{\bf 62}, 044022 (2000).
\bibitem{cc1999}
B.J.~Carr and A.A.~Coley,
Class. Quantum Grav
{\bf 16}, R31 (1999).
\bibitem{cc2000a}
  B.J.~Carr and A.A.~Coley,
  Phys. Rev. D
  {\bf 62}, 044023 (2000a).
\bibitem{cc2000b}
  B.J.~Carr and A.A.~Coley,
  Class. Quantum Grav. 
  {\bf 17}, 4339 (2000b).
\bibitem{ccgnu2000}
B.J.~Carr, A.A.~Coley, M.~Goliath, U.~Nilsson and C.~Uggla,
Phys. Rev. D {\bf 61}, 081502 (2000).
\bibitem{ccgnu2001}
B.J.~Carr, A.A.~Coley, M.~Goliath, U.~Nilsson and C.~Uggla,
Class. Quantum Grav. {\bf 18}, 303 (2001).
\bibitem{ch1989}
  B.~Carter and R.N.~Henriksen,
  Ann. Physique Supp.
  {\bf 14}, 47 (1989).
\bibitem{ch1991}
  B.~Carter and R.N.~Henriksen,
  J. Math. Phys.
  {\bf 32}, 2580 (1991).
\bibitem{choptuik1993}
M.W.~Choptuik,
Phys. Rev. Lett.
{\bf 70}, 9 (1993).
\bibitem{coley1997}
  A.A.~Coley,
  Class. Quantum Grav.
  {\bf 14}, 87 (1997).
\bibitem{ec1994}
C.R.~Evans and J.S.~Coleman,
Phys. Rev. Lett.
{\bf 72}, 1782 (1994).
\bibitem{fh1993}
  T.~Foglizzo and R.N.~Henriksen,
  Phys. Rev. D {\bf 48}, 4645 (1993).
\bibitem{fc1993}
  P.N.~Foster and R.A.~Chevalier,
  Astrophys. J.
  {\bf 416}, 303 (1993).
\bibitem{gw1980}
  P.~Goldreich and S.V.~Weber,
  Astrophys. J.
  {\bf 238}, 991 (1980).
\bibitem{gnu1998a}
M.~Goliath, U.~Nilsson and C.~Uggla,
Class. Quantum Grav. {\bf 15}, 167 (1998a).
\bibitem{gnu1998b}
M.~Goliath, U.~Nilsson and C.~Uggla,
Class. Quantum Grav. {\bf 15}, 2841 (1998b).
\bibitem{gundlach1998}
C.~Gundlach, 
Adv.~Theor.~Math.~Phys. {\bf 2}, 1 (1998).
\bibitem{gundlach1999} 
C.~Gundlach,
Living Rev. Rel. 
{\bf 2}, 4 (1999).
\bibitem{gundlach2002}
C.~Gundlach,
Phys. Rev. D {\bf 65}, 084021 (2002).
\bibitem{hm2000a}
T.~Hanawa and T.~Matsumoto,
Publ. Astron. Soc. Jpn.
{\bf 52}, 241 (2000a).
\bibitem{hm2000b}
T.~Hanawa and T.~Matsumoto,
Astrophys. J. 
{\bf 521}, 703 (2000b).
\bibitem{hn1997}
T.~Hanawa and K.~Nakayama,
Astrophys. J. 
{\bf 484}, 238 (1997).
\bibitem{harada1998}
T.~Harada,
Phys. Rev. D 
{\bf 58}, 104015 (1998).
\bibitem{harada2001}
T.~Harada,
Class. Quantum Grav. 
{\bf 18}, 4549 (2001).
\bibitem{hin2002}
T.~Harada, H.~Iguchi and K.~Nakao,
Prog. Theor. Phys. 
{\bf 107}, 449 (2002).
\bibitem{hm2001}
T.~Harada and H.~Maeda,
Phys. Rev. D 
{\bf 63}, 084022 (2001).
\bibitem{hms2003}
T.~Harada, H.~Maeda and B.~Semelin,
submitted to Phys. Rev. D,
gr-qc/0210027 (2003). 
\bibitem{hunter1977}
C.~Hunter,
Astrophys. J.
{\bf 218}, 834 (1977).
\bibitem{kha1995}
T.~Koike, T.~Hara and S.~Adachi,
Phys. Rev. Lett.
{\bf 74}, 5170 (1995).
\bibitem{kha1999}
T.~Koike, T.~Hara and S.~Adachi,
Phys. Rev. D 
{\bf 59}, 104008 (1999).
\bibitem{larson1969}
  R.B.~Larson,
  Mon. Not. R. Astr. Soc.
  {\bf 145}, 271 (1969).
\bibitem{madsen1988}
  M.S.~Madsen,
  Class. Quantum Grav. {\bf 5}, 627 (1988).
\bibitem{mh2001}
H.~Maeda and T.~Harada,
Phys. Rev. D 
{\bf 64}, 124024 (2001). 
\bibitem{mhio2002a}
H.~Maeda, T.~Harada, H.~Iguchi and N.~Okuyama, 
Phys. Rev. D {\bf 66}, 027501 (2002a).
\bibitem{mhio2002b}
H.~Maeda, T.~Harada, H.~Iguchi and N.~Okuyama, 
Prog. Theor. Phys. {\bf 108}, 819 (2002b).
\bibitem{mhio2003}
H.~Maeda, T.~Harada, H.~Iguchi and N.~Okuyama, 
in preparation (2003).
\bibitem{maison1996}
D.~Maison,
Phys. Lett. B
{\bf 366}, 82 (1996).
\bibitem{nc2000b}
D.W.~Neilsen and M.W.~Choptuik,
Class. Quantum Grav. {\bf 17}, 761 (2000).
\bibitem{op1987}
A.~Ori and T.~Piran,
Phys. Rev. Lett.
{\bf 59}, 2137 (1987).
\bibitem{op1988}
A.~Ori and T.~Piran,
Mon. Not. R. Astron. Soc.  
{\bf 234}, 821 (1988).
\bibitem{op1990}
A.~Ori and T.~Piran,
Phys. Rev. D
{\bf 42}, 1068 (1990).
\bibitem{penrose1969}
R.~Penrose,
Riv. Nuovo Cim. {\bf 1}, 252 (1969).
\bibitem{penrose1979}
R. Penrose, 
in {\it General Relativity, an Einstein Centenary Survey}, 
edited by S.W. Hawking and W. Israel (Cambridge University Press, 
Cambridge, England, 1979), p.581.
\bibitem{penston1969}
M.V.~Penston,
Mon. Not. R. Astron. Soc. {\bf 144}, 425 (1969).
\bibitem{shu1977ws1985}
  F.H.~Shu,
  Astrophys. J.
  {\bf 214}, 488 (1977).
\bibitem{ss1988}
  Y.~Suto and J.~Silk,
  Astrophys. J.
  {\bf 326}, 527 (1988).
\bibitem{ti1999}
T.~Tsuribe and S.~Inutsuka,
Astrophys. J. {\bf 526}, 307 (1999).
\bibitem{ws1985}
A.~Whitworth and D.~Summers,
Mon. Not. R. Astron. Soc. {\bf 214}, 1 (1985).
\bibitem{yahil1983}
  A.~Yahil,
  Astrophys. J.
  {\bf 265}, 1047 (1983).
\end{thebibliography}
\end{document}